\documentstyle[11pt]{article}

\begin{document} 
\centerline{\bf Numerical simulation of Quantum Teleportation in a chain of three nuclear spins system }
\centerline{\bf taking into account second neighbor iteration}
\vskip2pc
\centerline{G.V. L\'opez and L. Lara}
\centerline{Departamento de F\'{\i}sica, Universidad de Guadalajara}
\centerline{Apartado Postal  4-137, 44410 Guadalajara, Jalisco, M\'exico}
\vskip2pc
\centerline{PACS: 03.67.-a, 03.67.Lx, 03.67.Dd, 03.67.Hk}
\vskip2cm
\centerline{ABSTRACT}
\vskip1pc\noindent
For a one-dimensional chain of three nuclear spins (one half), we make the numerical simulation of
quantum teleportation of a given state from one end of the chain to the other end, taking into account
first and second neighbor interactions among the spins. It is shown that a well defined teleportation
protocol is achieved for a ratio of the first to second neighbor interaction coupling constant of
$J'/J\ge 0.04$. We also show that the optimum Rabi's frequency to control the non-resonant effects is
dominated by the second neighbor  interaction coupling parameter ($J'$).  
\vfil\eject\noindent 
{\bf 1. Introduction}
\vskip0.5pc\noindent
Quantum teleportation is a technique which allows us to move a quantum state of our quantum system from
one location to another location using quantum computation or quantum information [1]. Quantum
computation or quantum information uses quantum bits (called qubits) to handle the information on the
quantum system. A qubit is a superposition of any two quantum basic states of the system (called
$|0\rangle$ and $|1\rangle$), $\Psi=C_0|0\rangle+C_1|1\rangle$, such that $|C_0|^2+|C_1|^2=1$. One may
call $|0\rangle$ and $|1\rangle$ the basic qubits of the system. The $L$-tensorial product of $L$-basic
qubits form an $L$-register of $L$-qubits, $|x\rangle=|i_{L-1},\dots,i_0\rangle$ with $i_j=0,1$ for
$j=0,\dots,L-1$ ("0" for ground state and "1" for exited state). The set of these $L$-registers makes up
the $2^L$-dimensional Hilbert space where the quantum computer and quantum information work. A typical
element of this space is $\Psi=\sum C_x|x\rangle$, where $\sum |C_x|^2=1$, and $|C_x|^2$ gives us the
probability of having the state $|x\rangle$ after measurement. We are interested in studying the
phenomena of quantum teleportation in a solid state quantum computer formed by a one-dimensional chain
of nuclear spins [2], where a preliminary study of this technique was done using two-qubit registers and
considering first neighbor interaction between spins [3]. In this paper, we will consider also the second
neighbor interaction among the spins in a chain of three nuclear spins system. Thus, the phenomena of
quantum teleportation  from one end of the chain of spins to the other end is studied numerically.
Adding a quantum state (qubit) to our chain of three nuclear spins means to add another nuclear spin at
one end, having, then, a quantum computer working with four-qubit registers. We study the well
performance of the quantum teleportation algorithm  through the fidelity parameter and determine the
minimum value of the second neighbor interaction coupling constant to do this. Finally, we also see the
modification that the so called $2\pi k$-method could have with the consideration of second neighbor
iteration and its implication in our simulation
\vskip2pc
\leftline{\bf 2. Equation of Motion}
\vskip1pc\noindent
Consider a one-dimensional chain of four equally spaced nuclear-spins system (spin one half) making an
angle $\cos\theta=1/\sqrt{3}$ with respect the z-component of the magnetic field (chosen in this way to
kill the dipole-dipole interaction between spins) and having an rf-magnetic field in the transversal
plane. The magnetic field is given by
$${\bf B}=(b\cos(\omega t+\varphi), -b \sin(\omega t+\varphi), B(z))\ ,\eqno(1)$$
where $b$, $\omega$ and $\varphi$  are the amplitude, the angular frequency and the phase of the rf-field,
which could be different for different pulses. $B(z)$ is the amplitude of the z-component of the magnetic
field. Thus,
the Hamiltonian of the system up to second neighbor interaction is given by
$$H=-\sum_{k=0}^3{\bf \mu_k}\cdot {\bf B_k}-2J\hbar\sum_{k=0}^2I_k^zI_{k+1}^z
-2J'\hbar\sum_{k=0}^1I_k^zI_{k+2}^z\ ,\eqno(2)$$
where ${\bf\mu_k}$ represents the magnetic moment of the kth-nucleus which is given in terms of the
nuclear spin as ${\bf\mu_k}=\hbar\gamma(I_k^x,I_k^y, I_k^z)$, being $\gamma$ the proton gyromagnetic
ratio. ${\bf B_k}$ represents the magnetic field at the location of the $kth$-spin. The second term at the
right side of (2)  represents the  first neighbor spin interaction, and the third term represents the
second neighbor spin interaction. $J$ and $J'$ are the coupling constants for these interactions. This
Hamiltonian can be written in the  following way
$$H=H_0+W\ ,\eqno(3a)$$
where $H_0$ and $W$ are given by
$$H_0=-\hbar\left\{\sum_{k=0}^3\omega_kI_k^z+2J(I_0^zI_1^z+I_1^zI_2^z+I_2^zI_3^z)
+2J'(I_0^zI_2^z+I_1^zI_3^z)\right\}\eqno(3b)$$
and
$$W=-{\hbar\Omega\over 2}\sum_{k=0}^3\biggl[e^{i\omega t}I_k^++e^{-i\omega t}I_k^-\biggr]\ ,\eqno(3c)$$
where $\omega_k=\gamma B(z_k)$ is the Larmore frequency of the kth-spin, $\Omega=\gamma b$ is the Rabi's
frequency, and $I_k^{\pm}=I_k^x\pm iI_k^y$ represents the ascend operator (+) or the descend operator (-).
The Hamiltonian $H_0$ is diagonal on the basis $\{|i_3i_2i_1i_0\rangle\}$, where $i_j=0,1$ (zero for the
ground state and one for the exited state),
$$H_0|i_3i_2i_1i_0\rangle=E_{i_3i_2i_1i_0}|i_3i_2i_1i_0\rangle\ .\eqno(4a)$$
The eigenvalues $E_{i_3i_2i_1i_0}$ are given by
$$E_{i_3i_2i_1i_0}=-{\hbar\over 2}\biggl\{\sum_{k=0}^3(_1)^{i_k}\omega_k+J\sum_{k=0}^2(-1)^{i_k+i_{k+1}}
+J'\sum_{k=0}^1(-1)^{i_k+i_{k+2}}\biggr\}\ .\eqno(4b)$$
The term (3c) of the Hamiltonian allows to have a single spin transitions on the above eigenstates
by choosing the proper resonant frequency. 

\vskip1pc\noindent
To solve the Schr\"odinger equation
$$i\hbar{\partial\Psi\over\partial t}=H\Psi\ ,\eqno(5)$$ 
let us propose a solution of the form
$$\Psi(t)=\sum_{k=0}^{15}C_k(t)|k\rangle\ ,\eqno(6)$$
where we have used decimal notation for the eigenstates in (4a), $H_0|k\rangle=E_k|k\rangle$.
Substituting (6) in (5), multiplying for the bra $\langle m|$, and using the orthogonality relation 
$\langle m|k\rangle=\delta_{mk}$, we get the following equation for the coefficients
$$i\hbar\dot C_m=E_mC_m+\sum_{k=0}^{15}C_k\langle m|W|k\rangle\ \ m=0,\dots,15.\eqno(7)$$
Now, using the following transformation
$$C_m=D_me^{-iE_m t/\hbar}\ ,\eqno(8)$$
the fast oscillation term $E_mC_m$ of Eq. (7) is removed (this is equivalent to going to the interaction
representation), and the following equation is gotten for the coefficients $D_m$
$$i\dot D_m={1\over\hbar}\sum_{k=0}^{15}W_{mk}D_ke^{i\omega_{mk}t}\ ,\eqno(9a)$$
where $W_{mk}$  denotes the matrix elements $\langle m|W|k\rangle$, and $\omega_{mk}$ are defined as
$$\omega_{mk}={E_m-E_k\over\hbar}\ .\eqno(9b)$$
Eq. (9a) represents a set  of sixteen real coupling ordinary differential equations which can be solved
numerically, and where $W_{mk}'s$ are given by 
$$(W)=-{\hbar\Omega\over 2}\times\{0,\ z,\hbox{"or"}\ z^*\}\ ,\eqno(9c)$$
where $z$ is defined as $z=e^{i(\omega t+\varphi)}$, and $z^*$ is its complex conjugated. 
\vskip0.5pc\noindent
A pulse of time length $\tau$, phase $\varphi$, and resonant frequency $\omega=\omega_{i,j}$ will be
denoted by $R_{i,j}(\Omega\tau,\varphi)$, and it is understood that such a pulse implies the solution of
Eq. (9a) with given initial conditions.
\vskip2pc
\leftline{\bf 3. Quantum teleportation and numerical simulation}
\vskip1pc\noindent
The basic idea of quantum teleportation [1] is that Alice (left end qubit in our chain of three qubits)
and Bob (the other end qubit) are related through and entangled state,
$$\Psi_e={1\over\sqrt{2}}\biggl(|0_A00_B\rangle+|1_A01_B\rangle\biggr)\ .\eqno(10)$$
So, neglecting decoherence effects which could destroy this entangled state, we adjoin to Alice an
arbitrary state $\Psi_x$ ,
$$\Psi_x=C_0^x|0\rangle+C_1^x|1\rangle\ ,\eqno(11)$$
resulting the quantum state ($\Psi_1=\Psi_x\otimes\Psi_e$)
$$\Psi_1={1\over\sqrt{2}}\biggl[C_0^x|0000\rangle+C_0^x|0101\rangle+C_1^x|1000\rangle+C_1^x|1101\rangle\biggr]\
.\eqno(12)$$
Then, one applies a Controlled-Not (CN) operation in this added stated and Alice,
$$\widehat{ CN}_{23}|i_3,i_2,i_1,i_0\rangle=|i_3, i_2\oplus i_3, i_2, i_0\rangle\ ,\eqno(13)$$
where $i_2\oplus i_3=i_2+i_3 (mod~2)$, resulting the state ($\Psi_2=\widehat{CN}_{23}\Psi_1$)
$$\Psi_2={1\over\sqrt{2}}\biggl[C_0^x|0000\rangle+C_0^x|0101\rangle+C_1^x|1100\rangle+C_1^x|1001\rangle\biggr]\
.\eqno(14)$$
Finally, one applies a superposition (Hadamar) operation in the added state location,
$$A_3\cases{|0000\rangle\cr |1000\rangle}={1\over\sqrt{2}}\cases{|0000\rangle+|1000\rangle\cr
|0000\rangle-|1000\rangle}\ ,\eqno(15)$$
resulting, after some rearrangements, the state ($\Psi_3=A_3\Psi_2$)
\begin{eqnarray*}
\Psi_3&=&{1\over 2}\Biggl\{|00\rangle\otimes|0\rangle\biggl(C_0^x|0\rangle+C_1^x|1\rangle\biggr)+
|01\rangle\otimes|0\rangle\otimes\biggl(C_0^x|1\rangle+C_1^x|0\rangle\biggr)\\
& &+|10\rangle\otimes|0\rangle\otimes\biggl(C_0^x|0\rangle-C_1^x|1\rangle\biggr)+
|11\rangle\otimes|0\rangle\otimes\biggl(C_0^x|1\rangle-C_1^x|0\rangle\biggr)\Biggr\}\ .
\end{eqnarray*}
$$\eqno(16)$$
When Alice measures both of her qubits, there are four possible cases ($|00\rangle$, $|01\rangle$,
$|10\rangle$, and $|11\rangle$), and for each case Bob will get the original $\Psi_x$ state by applying
a proper operation to his state: identity, Not ($\widehat{ N}$), $\sigma_z$, or $\sigma_z\widehat N$,
where the action of these operators are as follows: $\widehat N|0\rangle=|1\rangle$, $\widehat
N|1\rangle=|0\rangle$,
$\sigma_z|0\rangle=|0\rangle$, and $\sigma_z|1\rangle=-|1\rangle$. The most important features in this
algorithm are that Alice and Bob do not need to know the state $\Psi_x$, and there are always
four possibilities for Alice measurement.
\vskip0.5pc\noindent
Now, the algorithm we need to implement this quantum teleportation technique  in our one-dimensional
three nuclear spins system varies a little from the scheme presented above since our three-qubits
quantum computer will start from the ground state
$$\Psi_{00}=|000\rangle\ .\eqno(17a)$$
Therefore, one needs first to attach  the unknown state (11) to (17a) to get the initial four-spins
wave function ($\Psi_0=\Psi_x\otimes\Psi_{00}$),
$$\Psi_0=C_0^x|0000\rangle+C_1^x|1000\rangle\ ,\eqno(17b)$$
where $|C_0^x|^2+|C_1^x|^2=1$. The entangled state (12) is obtained with the following three pulses
$$\Psi_1=R_{4,5}(\pi,-3\pi/2)R_{8,12}(\pi/2,-\pi/2)R_{0,4}(\pi/2,-\pi/2)\Psi_0\ .\eqno(16c)$$
The Controlled-Not operation $\widehat{CN}_{23}\Psi_1$ is gotten  through the following two pulses
$$\Psi_2=R_{9,13}(\pi,3\pi/2)R_{8,12}(\pi,-3\pi/2)\Psi_1\ .\eqno(17d)$$
Finally, the wave function (16) is gotten after the application of the following two pulses to the above
wave function
$$\Psi_3=R_{4,12}(\pi/2,-3\pi/2)R_{1,9}(\pi/2,-3\pi/2)\Psi_2\ .\eqno(17e)$$
\vskip0.5pc\noindent
To make the numerical simulation of this algorithm, we have chosen the following  parameters
in units of $2\pi\times$Mhz,
$$\omega_0=100\ ,\ \omega_1=200\ ,\ \omega_2=400\ ,\ \omega_3=800,\ J=10\ ,\ J'=0.4\ ,\
\Omega=0.1\ .\eqno(18a)$$ 
These parameters were chosen in this way to have a clear definition on our Zeeman spectrum  and
transitions among them. Of course, our main results are applied to the actual current design parameters
of reference [4]. On the other hand, we selected the state (11) with the following coefficients
$$C_0^x={1\over 3}\hskip3pc\hbox{and}\hskip3pc C_1^x={\sqrt{8}\over 3}\ .\eqno(18b)$$
Fig. 1 shows the Zeeman spectrum of the four nuclear spins system with the transitions used during our
teleportation simulation. Fig. 2a shows the probabilities $|C_0|^2$, $|C_5|^2$, $|C_8|^2$ and
$|C_{13}|^2$ during the first three pulses, where the formation of the entangled wave function (12) is
shown, from the initial state (17b). Fig. 2b shows the probabilities $|C_0|^2$, $|C_5|^2$, $|C_9|^2$ and
$|C_{12}|^2$ during the following two pulses to get at the end the wave function (14). Fig. 2c shows the
probabilities $|C_0|^2$, $|C_8|^2$, $|C_5|^2$, $|C_{13}|^2$, $|C_{12}|^2$, $|C_4|^2$, $|C_9|^2$ and
$|C_1|^2$ during the last two pulses to get at the end the desired function (16). Note that at the end
of our algorithm one expects the following values for these probabilities 
$$|C_0|^2=|C_8|^2=|C_5|^2=|C_{13}|^2={|C_0^x|^2\over 4}\ ,\eqno(19a)$$
and
$$|C_{12}|^2=|C_4|^2=|C_9|^2=|C_{1}|^2={|C_1^x|^2\over 4}\eqno(19b)$$
To get a better
feeling what is going on on each process, we calculate the z-component of the expected values of the
spin for each qubit. These expected values are given by  
$$\langle I_0^z\rangle={1\over 2}\sum_{k=0}^{15} (-1)^k|C_k(t)|^2\ ,\eqno(20a)$$
\begin{eqnarray*}
\langle I_1^z\rangle&=& {1\over 2}\Biggl\{|C_0|^2+|C_1|^2-|C_2|^2-|C_3|^2+|C_4|^2+|C_5|^2-|C_6|^2-|C_7|^2
\\ \\
& &+|C_8|^2+|C_9|^2-|C_{10}|^2-|C_{11}|^2+|C_{12}|^2+|C_{13}|^2-|C_{14}|^2-|C_{15}|^2\Biggr\}\ ,
\end{eqnarray*}
$$\eqno(20b)$$
\begin{eqnarray*}
\langle I_2^z\rangle&=& {1\over 2}\Biggl\{|C_0|^2+|C_1|^2+|C_2|^2+|C_3|^2-|C_4|^2-|C_5|^2-|C_6|^2-|C_7|^2
\\ \\
& &+|C_8|^2+|C_9|^2+|C_{10}|^2+|C_{11}|^2-|C_{12}|^2-|C_{13}|^2-|C_{14}|^2-|C_{15}|^2\Biggr\}\ ,
\end{eqnarray*}
$$\eqno(20c)$$
and
$$\langle I_3^z\rangle={1\over 2}\sum_{k=0}^7|C_k|^2-{1\over 2}\sum_{k=8}^{15}|C_k|^2\ .\eqno(20d)$$
Figs. 3a, 3b, and 3c show these expected values during entangled formation, controlled-not operation, and
final teleportation. As one can see, these behavior is what one could expected for each process.
Fig. 4a shows the probabilities (19a) and (19b) at the end of the teleportation (wave function (16)), and
Fig. 4b shows the probabilities of the non-resonant states involved in the dynamics. 
\vskip0.5pc\noindent
To see the values of
the second neighbor interaction coupling parameter from which one could have a well defined
teleportation algorithm, we study the fidelity parameter [5],
$$F=\langle\Psi_{expected}|\Psi\rangle\ ,\eqno(21)$$
where $\Psi_{expected}$ is the ideal wave function (16). Fig. 5 shows this fidelity parameter as a
function of the ratio of second to first neighbor coupling parameters ($J'/J$). As one can see, a well
defined teleportation algorithm is gotten for $J'/J\ge 0.04$.
\vskip1pc\noindent
\leftline{\bf 4. Quantum teleportation and the $2\pi k$-method}
\vskip0.5pc\noindent
One of the important results from the consideration of first neighbor interaction and the selection of
the parameter as $J/\Delta\omega\ll 1$ and $\Omega/\Delta\omega\ll 1$ is the possibility to choose the
Rabi's frequency
$\Omega$ in such a way that the non-resonant effects are eliminated. This procedure is called the $2\pi
k$-method [4], and this Rabi's frequency is chosen as $\Omega=|\Delta|/\sqrt{4k^2-1}$ for a $\pi$-pulse,
where $k$ is an integer number, $\Delta$ is the detuning parameter ($\Delta=(E_p-E_m)/\hbar-\omega$)
between the states $|p\rangle$ and $|m\rangle$ when the resonant frequency is $\omega$, and this detuning
parameter is proportional to the first neighbor coupling constant $J$. Let us see how
this detuning parameter could be modified due to second neighbor interaction. Assuming that the states
$|p\rangle$ and
$|m\rangle$ are are the only ones involved in the dynamics, from Eq. (9a), one has
$$i\dot D_m={W_{mp}\over\hbar}D_pe^{i\omega_{mp}t}\ ,\hskip0.5pc\hbox{and}\hskip1.5pc i\dot
D_p={W_{pm}\over\hbar}D_m e^{i\omega_{pm}t}\ .\eqno(22)$$
Thus, given the initial conditions $D_p(0)=C_p(0)$ and $D_m(0)=C_m(0)$, the solution is readily given by
$$D_p(t)=\Biggl\{C_p(0)\biggl[\cos{\Omega_et\over 2}-i{\Delta\over\Omega_e}\sin{\Omega_et\over 2}\biggr]
+i{\Omega C_m(0)\over\Omega_e}\sin{\Omega_e t\over 2}\Biggr\}e^{i\Delta t\over 2}\eqno(23a)$$
and
$$D_m(t)=\Biggl\{C_m(0)\biggl[\cos{\Omega_et\over 2}-i{\Delta\over\Omega_e}\sin{\Omega_et\over 2}\biggr]
+i{\Omega C_p(0)\over\Omega_e}\sin{\Omega_e t\over 2}\Biggr\}e^{-i\Delta t\over 2}\ ,\eqno(23b)$$
where $\Omega_e$ is defined as $\Omega_e=\sqrt{\Omega^2+\Delta^2}$. For a
$\pi$-pulse ($t=\tau=\pi/\Omega$), one can select the term $\Omega_e\pi/2\Omega$ to be equal to any
multiple of $\pi$,  $\Omega_e\pi/2\Omega=k\pi$, to get the condition
$\Omega=|\Delta|/\sqrt{4k^2-1}$. this condition gets rid of the non-resonant terms since from 
Eqs. (20a) and (20b) one gets
$$D_p(\tau)=(-1)^kC_p(0)e^{i\Delta\pi/2\Omega}\ ,\hskip0.5pc\hbox{and}\hskip0.5pc 
D_m(\tau)=(-1)^kC_m(0)e^{-i\Delta\pi/2\Omega}\ .$$
For a $\pi/2$-pulse ($t=\tau=\pi/2\Omega$), one can select the term $\Omega_e\pi/4\Omega$ to be equal to
any multiple of $\pi$,  $\Omega_e\pi/4\Omega=k\pi$, to get the condition
$\Omega=|\Delta|/\sqrt{16k^2-1}$. this condition gets rid also of the non-resonant terms.
\vskip0.5pc\noindent
Now, if for example one selects a resonant transition which contains the Larmore frequency $\omega_0$,
these frequencies could be $\omega_0+J+J'$, $\omega_0-J+J'$, $\omega_0-J-J'$ or $\omega_0+J-J'$ which
correspond to the transitions (decimal notation) $|0\rangle\leftrightarrow |1\rangle~(|10\rangle
\leftrightarrow |11\rangle)$, $|2\rangle\leftrightarrow|3\rangle$, $|6\rangle\leftrightarrow|7\rangle$, 
and $|2\rangle\leftrightarrow|3\rangle$. So, all of these states are pertubated, and the frequency
difference $\Delta$ may have the values $2J$, $2J'$, $2J+2J'$, or $2J-2J'$. For other Larmore frequencies
the additional values of the detuning parameter are $4J$, and $4J+2J'$. Thus, let us denote by
$\Omega_{\Delta}^{(k)}$ the Rabi's frequency selected by this method,
$$\Omega_{\Delta}^{(k)}={|\Delta|\over\sqrt{4 k^2-1}}\ ,\hskip2pc\hbox{$\pi$-pulse}\eqno(24a)$$
and
$$\widetilde{\Omega}_{\Delta}^{(k)}={|\Delta|\over\sqrt{16 k^2-1}}\
,\hskip2pc\hbox{$\pi/2$-pulse}\eqno(24b)$$ where $\Delta$ can have the values $4J+2J'$, $4J$, $2J+2J'$,
$2J$ or $2J'$. To see the dependence of our teleportation algorithm with respect the Rabi's
frequency, we use again the fidelity parameter parameter (21). With the same values for our parameters
as (18a) but $\Omega$, Fig. 6 shows the fidelity parameter as a function of the Rabi's frequency. Dashed
vertical lines mark the omega values where the peaks ocurres. These peaks correspond to some specific
omega defined through the $2\pi k$-method. For example, the line (1), (2) and (3) correspond to the
following Rabi's frequency values  
$$\Omega_{4J+2J'}^{(202)}\approx
\Omega_{4J}^{(199)}\approx
\Omega_{2J+2J'}^{(103)}\approx
\Omega_{2J}^{(100)}\approx
\Omega_{2J'}^{(4)}=0.10079\ ,$$

$$\Omega_{4J+2J'}^{(305)}\approx
\Omega_{4J}^{(304)}\approx
\Omega_{2J+2J'}^{(156)}\approx
\Omega_{2J}^{(150)}\approx
\Omega_{2J'}^{(6)}=0.066889\ ,$$
and
$$\Omega_{4J+2J'}^{(407)}\approx
\Omega_{4J}^{(400)}\approx
\Omega_{2J+2J'}^{(208)}\approx
\Omega_{2J}^{(200)}\approx
\Omega_{2J'}^{(8)}=0.050098\ .$$
As one can see in Fig. 7a, where we have plotted $\Omega_{\Delta}^{(k)}$ (for the detuning values
mentioned above) and where the corresponding dashed vertical lines of Fig. 6  have been drawn, around
these lines there are several other values of $\Omega_{4J+2J'}^{(k)}$, $\Omega_{4J}^{(k)}$,
$\Omega_{2J+2J'}^{(k)}$ and 
$\Omega_{2J}^{(k)}$ which, in principle, should cause a peak in the fidelity parameter (because they
belong to the $2\pi k$-method). However, they do not appear at all on Fig. 6. This means that the peaks
values on the fidelity parameter are fully dominated by the second neighbor coupling interaction parameter
($J'$). On the other hand, the reason why only even numbers of $k$ appears for these peaks for  
$\Omega_{2J'}^{(k)}$ can be seen in Fig. 7b where $\Omega_{2J'}^{(k)}$ and 
$\widetilde{\Omega}_{2J'}^{(k)}$ have been plotted as a function of $k$. On this plot one sees that 
the jth-dashed lines correspond to $\widetilde{\Omega}_{2J'}^{(j)}=\Omega_{2J'}^{2j}$. Therefore, the
peaks on fidelity correspond to $2\pi k$-method dominated by $J'$ and by $\pi/2$-pulses. This seems
reasonable since our teleportation algorithm starts with $\pi/2$-pulses and finishes with $\pi/2$-pulses.
\vfil\eject\vskip3pc\noindent
\leftline{\bf 5. Conclusions}
\vskip0.5pc\noindent
We have made a numerical simulation of teleportation in a solid
state quantum computer modeled by one-dimensional chain of three nuclear spins (one half), and considering
first and second neighbor interactions. We have shown that a good teleportation algorithm can be gotten
if the ratio of second to first neighbor interaction constants is chosen such that $J'/J\ge 0.04$. 
We also
studied the effect of the second neighbor interaction on the detuning parameter which is used in the so
called $2\pi k$-method to eliminate non-resonant transitions, and we have shown that the application of
this method in our teleportation algorithm is not so simple since the detuning parameter varies with both
parameters $J$ and $J'$ (first and second neighbor coupling interactions). However, the peaks on the
fidelity parameter are dominated by the second neighbor coupling parameter and the $\pi/2$-pulses.
\vskip5pc\noindent
\leftline{\bf Acknowledgements}
\vskip0.5pc\noindent
 This work was supported by SEP under the contract PROMEP/103.5/04/1911 and the University of Guadalajara.
\vfil\eject

\leftline{\bf Figure Captions}
\vskip1pc\noindent
Fig. 1 Energy levels and resonant frequencies used within the algorithm.
\vskip0.5pc\noindent
Fig. 2 Probabilities $|C_k|^2: (k)$. (a) Formation of the entangled state, wave
function (12).  (b) Formation of the wave function (14). (c) Formation the wave function (16).
\vskip0.5pc\noindent
Fig. 3 Expected values $\langle I_k^z\rangle$: (k=0,1,2,3). (a) During formation of wave function (12).
(b) During formation of the wave function (14). (c) During formation of the wave function (16). 
\vskip0.5pc\noindent
Fig. 4 Probabilities $|C_k|^2$. (a) For the expected registers $k=0,8,5,13,12,4,9,1$. (b) For the
non-resonant states $k=2,3,6,7,10,11,14,15$.
\vskip0.5pc\noindent
Fig. 5 Fidelity parameter as a function of $J'/J$.
\vskip0.5pc\noindent
Fig. 6 Fidelity parameter as a function of $\Omega$.
\vskip0.5pc\noindent
Fig. 7 (a): Rabi frequency $\Omega_{\Delta}^{(k)}$ as a function of $k$ for 
$\Delta=4J+2J'$ [1], $\Delta=4J$ [2], $\Delta=2J+2J'$ [3], $\Delta=2J$ [4], $\Delta=2J'$ [5]. Dashed lines
$(j)$  for $j=1,\dots,8$ correspond to Fig. 6. (b): Rabi frequency $\Omega_{\Delta}^{(k)}$ [1] and
$\widetilde{\Omega}_{\Delta}^{(k)}$ [2] as a function of $k$. Dashed lines
$(j)$ for $j=1,\dots,8$ correspond to Fig. 6.
\vfil\eject
\leftline{\bf References}
\obeylines{
1. C.H. Benneth, G. Brassard, C. Cr\'epeau, R. Jozsa, A. Peres, and W. Wootters
\quad Phys. Rev. Lett., {\bf 70} (1993) 1895.
2. S. Lloyd, Science, {\bf 261} (1995) 1569.
\quad S. Lloyd, Sci. Amer., {\bf 273} (1995)  140.
\quad G.P. Berman, G.D. Doolen, D.J. Kamenev, G.V. L\'opez, and V.I. Tsifrinovich
\quad Phys. Rev. A, {\bf 6106} (2000) 2305.
3. G.P. Berman, G.D. Doolen, G.V. L\'opez, and V.I. Tsifrinovich,
\quad quant-ph/9802015 (1998).
4. G.P. Berman, G.D. Doolen, D.J. Kamenev, G.V. L\'opez, and V.I. Tsifrinovich
\quad Contemporary Mathematics, {\bf 305} (2002) 13.
5. A. Peres, Phys. Rev. A {\bf 30} (1984) 1610.
\quad B. Schumacher, Phys. Rev. A.,{\bf 51} (1995) 2738.
}

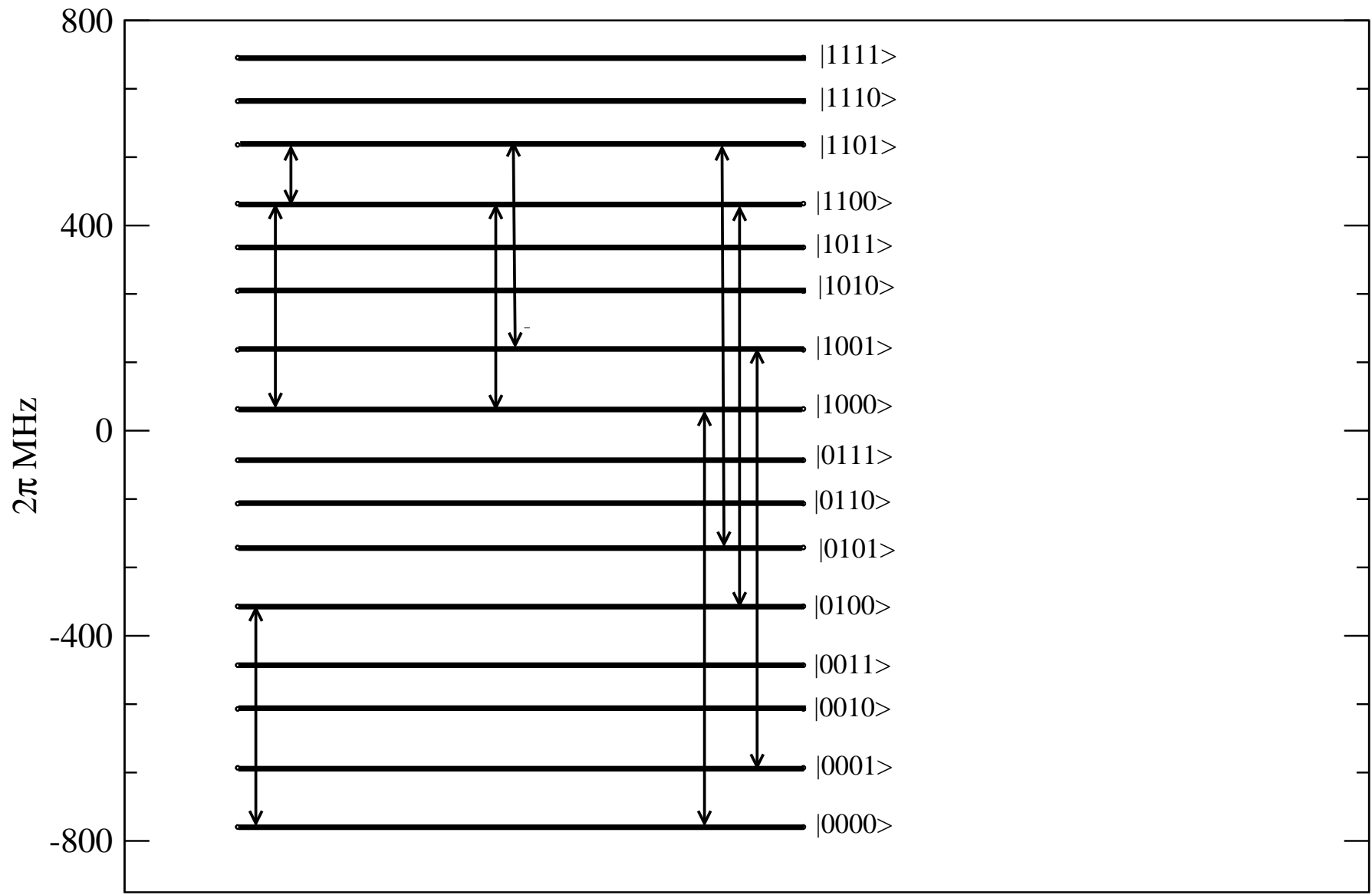
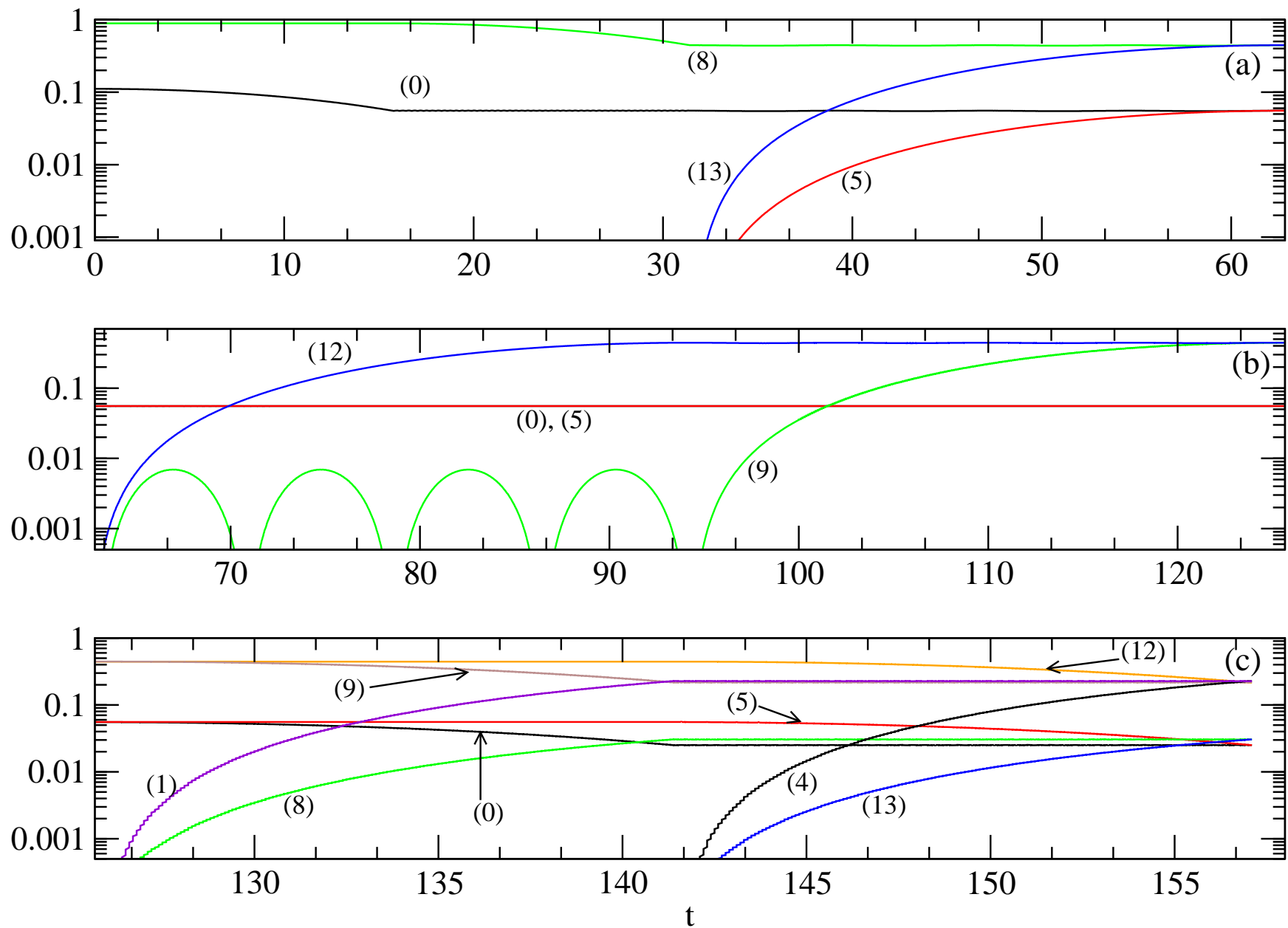
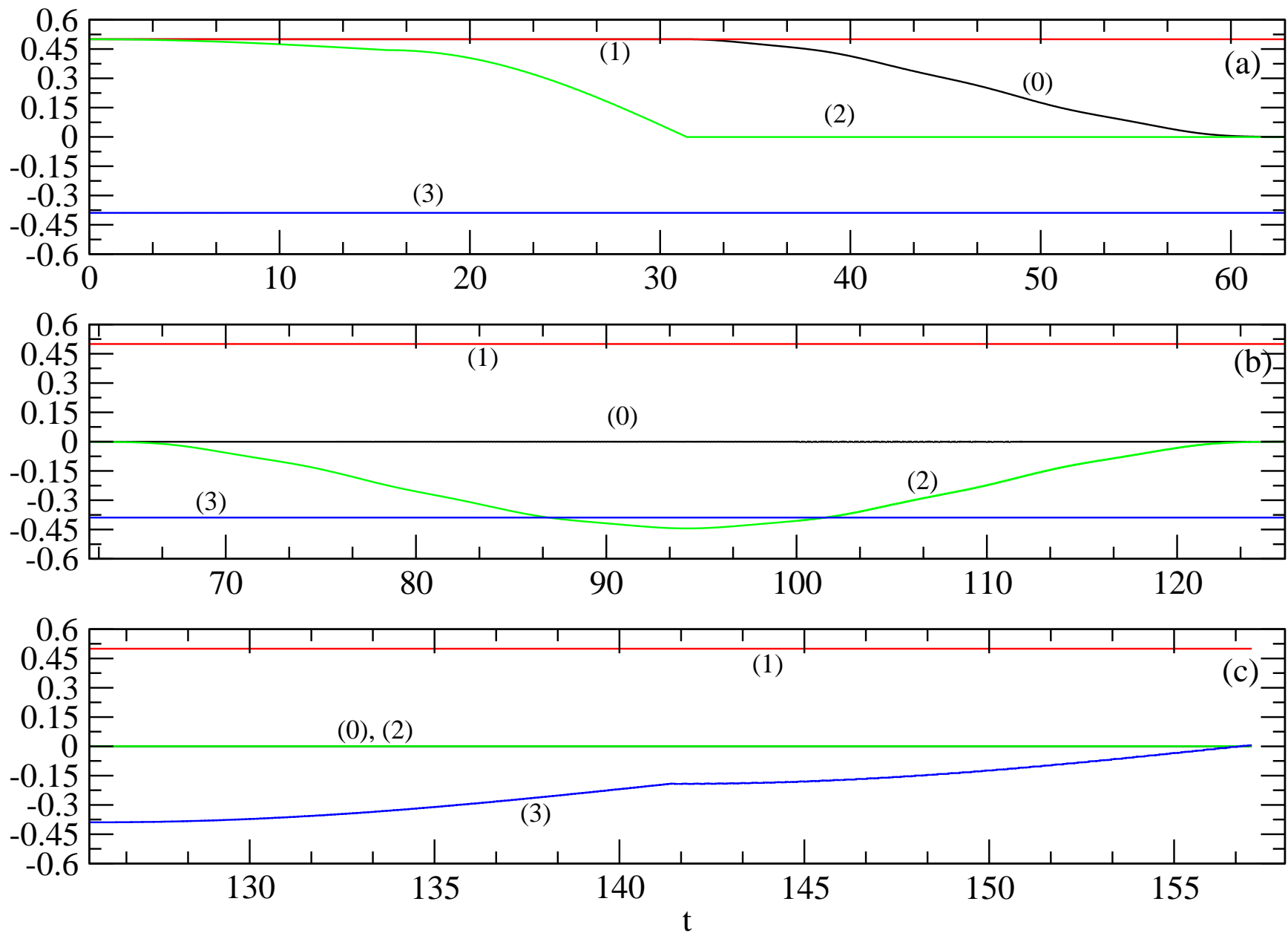
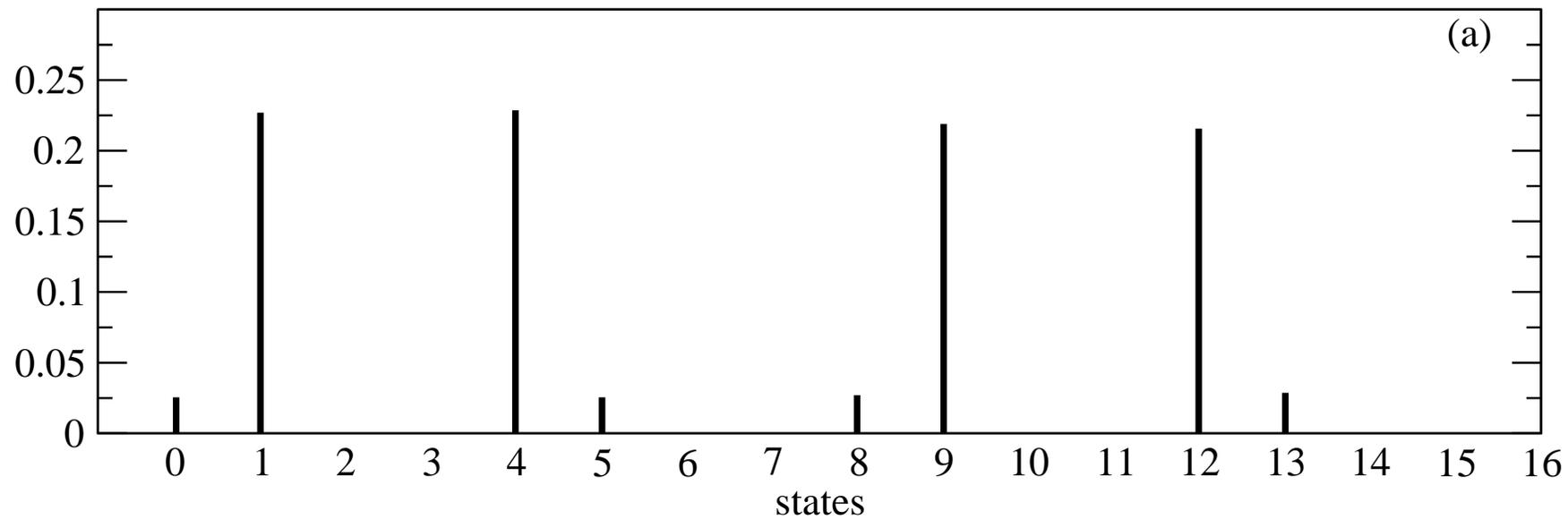
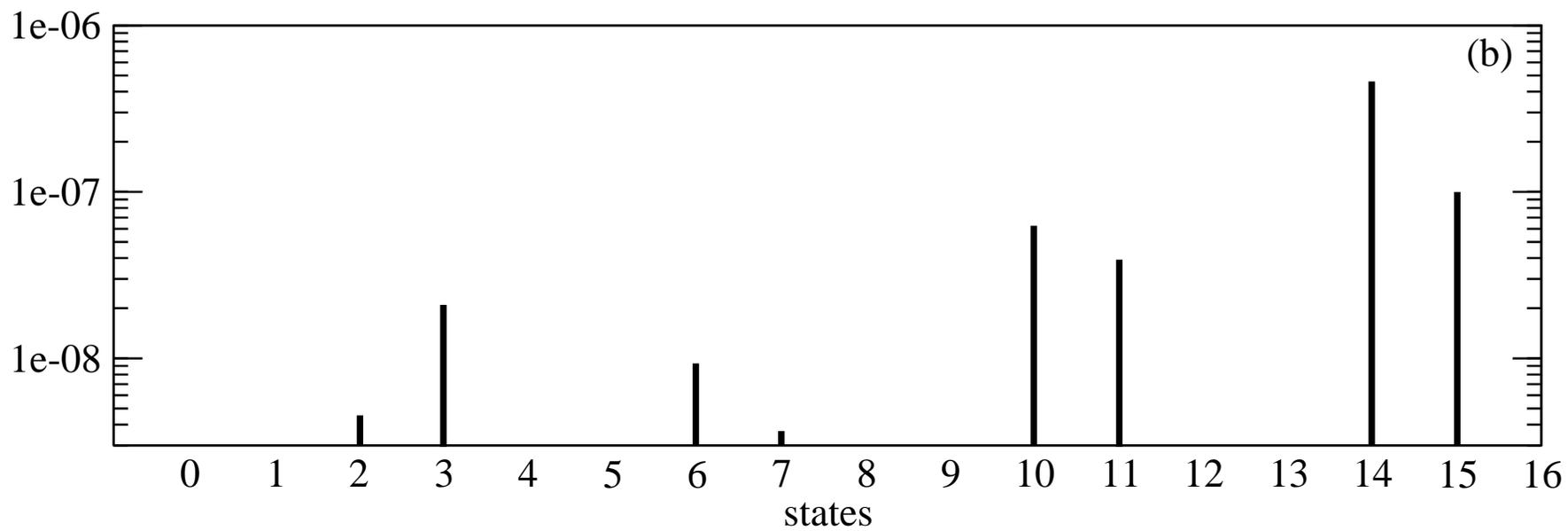
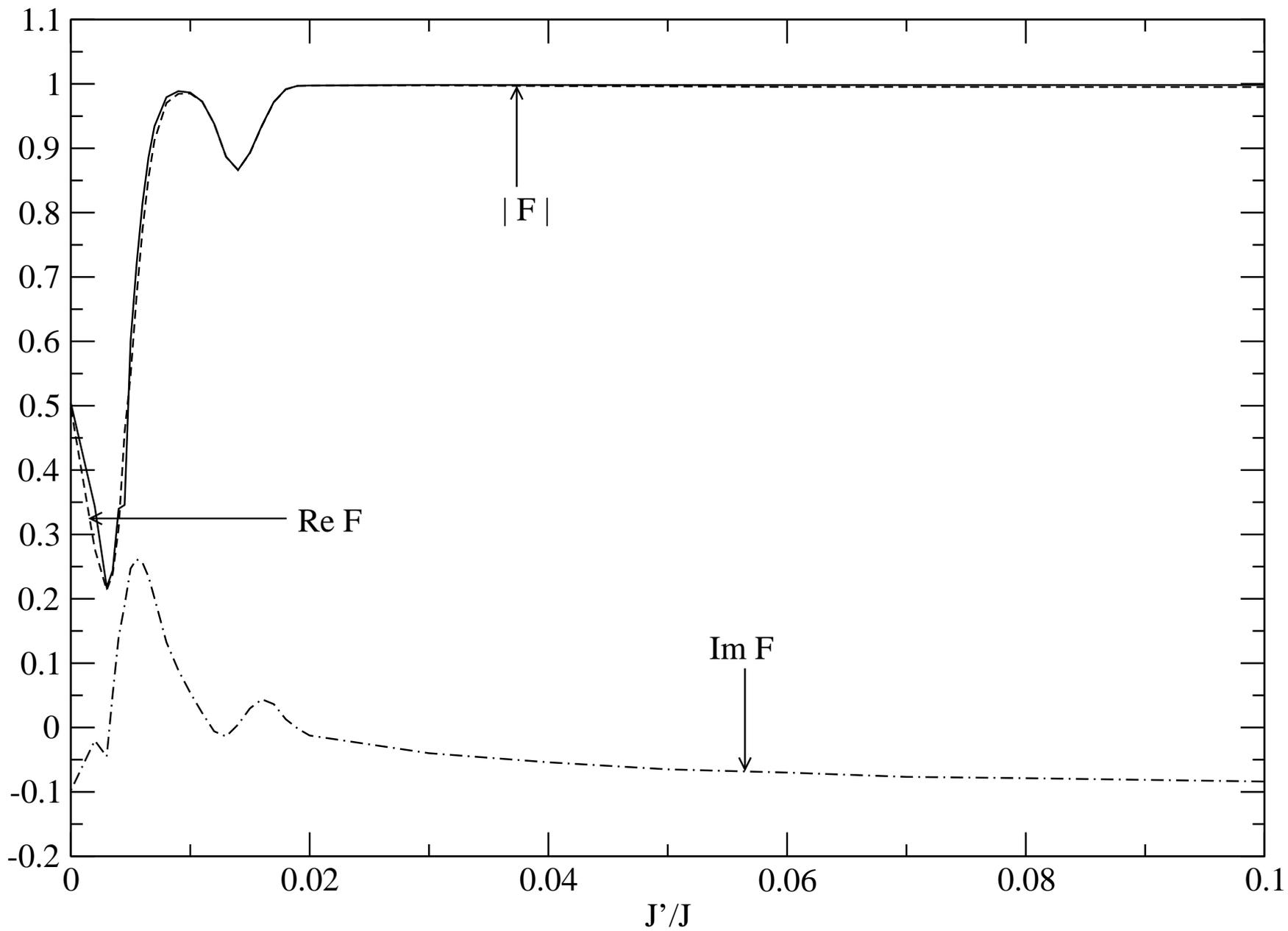
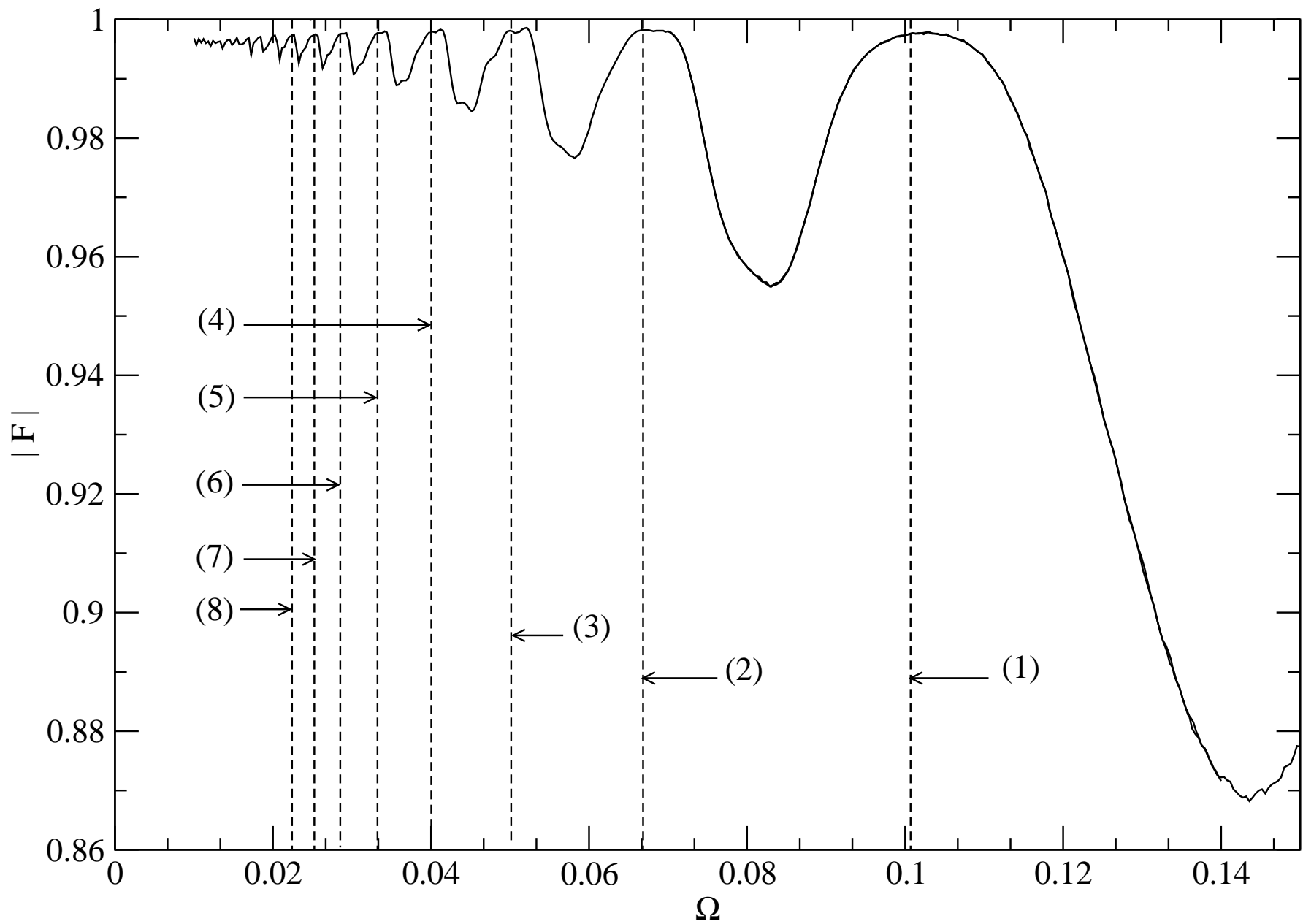
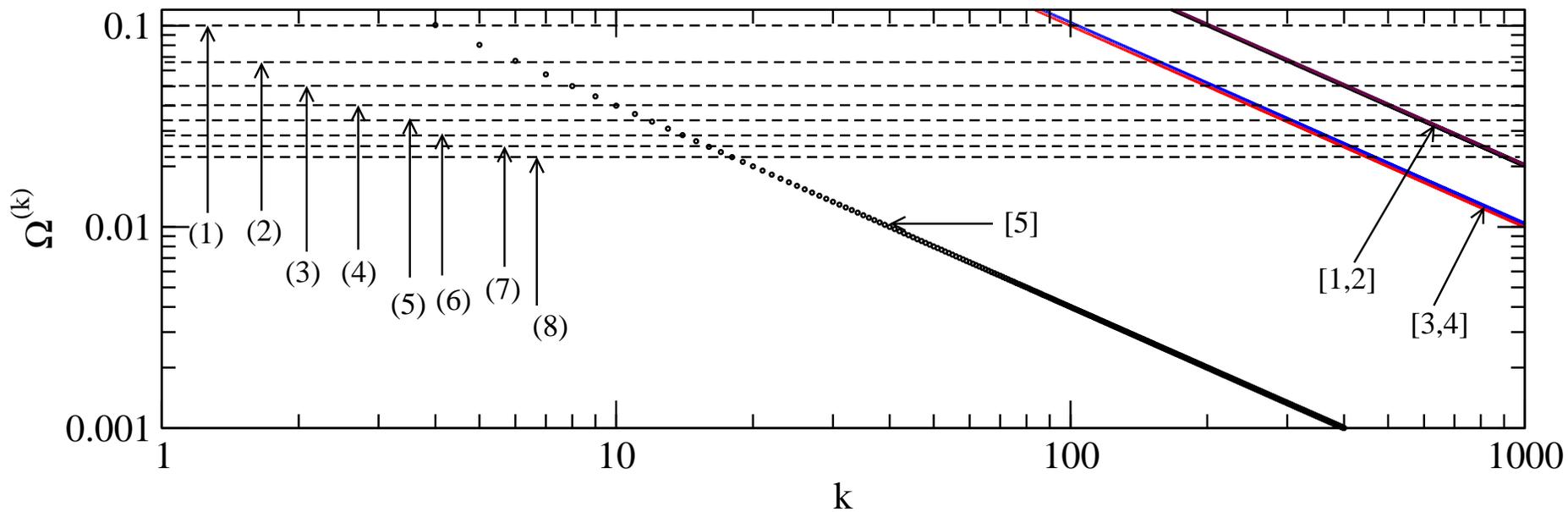
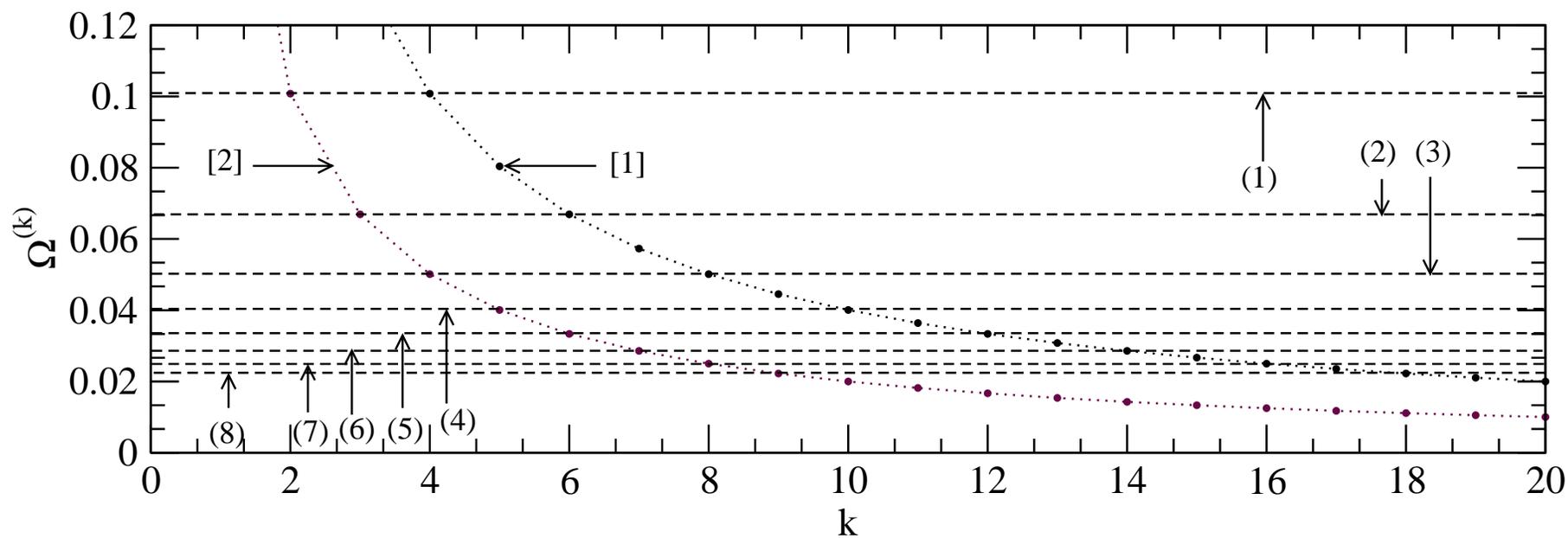
\end{document}